\documentclass[cmfonts]{aipproc}
\layoutstyle{6x9}

\newcommand{\Myr}{M$_{\odot}$~yr$^{-1}$}
\newcommand{\Ha}{H$\alpha$}
\newcommand{\Hb}{H$\beta$}
\newcommand{\Brg}{Br$\gamma$}

\newcommand{\Teff}{$T_{\rm{eff}}$}

\newcommand{\Ms}{M$_{\odot}$}

\newcommand{\kms}{km~s$^{-1}$}
\newcommand{\siiv}{Si~IV $\lambda$1400}
\newcommand{\civ}{C~IV $\lambda$1550}

\begin{document}

\title{Age-Dating of Starburst Galaxies}

\author{Claus Leitherer}{
  address={Space Telescope Science Institute, 3700 San Martin Drive, Baltimore, MD 21218}
}

\begin{abstract} This review covers age-dating methods applied to young stellar populations in starburst galaxies with ages of $10^8$~yr and less. First, recent advances in stellar modeling, both for the interior and the atmospheres, are discussed and spectral lines and colors suitable for age-dating are identified. These age indicators are then applied to single stellar populations, the simplest building blocks of starbursts. More complex systems, such as luminous starburst galaxies, are a challenge for chronometric measurements. The dust reddening, metallicity, stellar initial mass function, and inhomogeneities of the interstellar medium introduce degeneracies which are often hard to resolve. Finally, some shortcomings in the stellar models are highlighted.

\end{abstract}

\maketitle

%%%%%%%%%%%%%%%%%%%%%%%%%%%%%%%%%%%%%%%%%%%%
%% MAINMATTER
%%%%%%%%%%%%%%%%%%%%%%%%%%%%%%%%%%%%%%%%%%%%

\section{Introduction}

The evolution of normal galaxies is predominantly driven by stars. Young, massive stars with masses above $\sim$5~\Ms\ forming in starbursts provide the thermal and non-thermal luminosity responsible for the observed global galactic parameters (Heckman 1998). The tight connection of stellar and galactic properties permits the use of stars as tracers of starbursts as a whole. Therefore, understanding and reading the clock set by stellar evolution is a necessary prerequisite for age-dating starbursts. It is not a sufficient condition, however, as the complex morphology of stars, gas, and dust in a starburst make it difficult to interpret such an idealized clock.

\section{Stellar Models}

The stellar clock is driven by the interior evolution, whose main determinants are the chemical composition, stellar mass, and the decline of mass with time. In addition, massive-star evolution depends rather critically on mixing processes which determine the relative efficiency of the convective vs. radiative energy transport. Since convection is the more efficient process, any mechanism in its favor (such as rotation), will increase the stellar temperatures in the Hertzsprung-Russell diagram (HRD) in comparison with a radiative model (Yi 2003). Despite remaining uncertainties, the overall evolution in the upper HRD is fairly well understood (Chiosi \& Maeder 1986; Maeder \& Meynet 2000).

Massive stars follow one of two major evolutionary channels. If their initial masses are below $\sim$25~\Ms\ (for solar chemical composition), the core-contraction is mirrored by an envelope expansion until they reach the Hayashi line in the red. Observationally, this evolution is identified as the sequence OB star $\rightarrow$ blue supergiant $\rightarrow$ red supergiant (RSG). More massive stars experience higher mass loss because of the strong luminosity dependence of radiatively driven winds (Kudritzki \& Puls 2000). As a result, the He-rich cores are exposed before the RSG phase is reached, and the corresponding hotter surface temperatures lead to a reversal of the evolution back to the blue part of the HRD. The spectroscopic phases of this sequence are OB star $\rightarrow$ blue supergiant $\rightarrow$ Wolf-Rayet (W-R) star. A W-R population indicates younger, more massive stars than a RSG population.

Calibrating and verifying these predictions is non-trivial because of the scarcity of massive stars, their uncertain distances, and their generally elusive parameters. Therefore, the recent mass determination of the eclipsing spectroscopic \mbox{W-R} binary WR20a from an orbit analysis by Bonanos et al. (2004) and Rauw et al. (2004) is significant: the two components have masses of $83 \pm 5$~\Ms\ and $82 \pm 5$~\Ms, making them the most massive stars with direct mass determinations. Their properties agree with the predictions of stellar evolution theory. 

Evolution models by themselves are of limited usefulness for comparison with observables because they lack detailed predictions of spectral features. Therefore they are usually linked to a spectral library, which can be either empirical or theoretical (e.g., Lejeune \& Fernandes 2002). Recently, there has been a shift in preference away from empirical to theoretical libraries. The main reasons are the reliability of the latest generation of model atmospheres and the need to cover the full parameter space which is otherwise observationally inaccessible. Galactic stars have the chemical evolutionary history imprinted, and their spectra are therefore badly suited for comparison with those of, e.g., elliptical (Thomas et al. 2004) or Lyman-break galaxies (Mehlert et al. 2002).   

\begin{figure}
  \includegraphics[height=0.6\columnwidth]{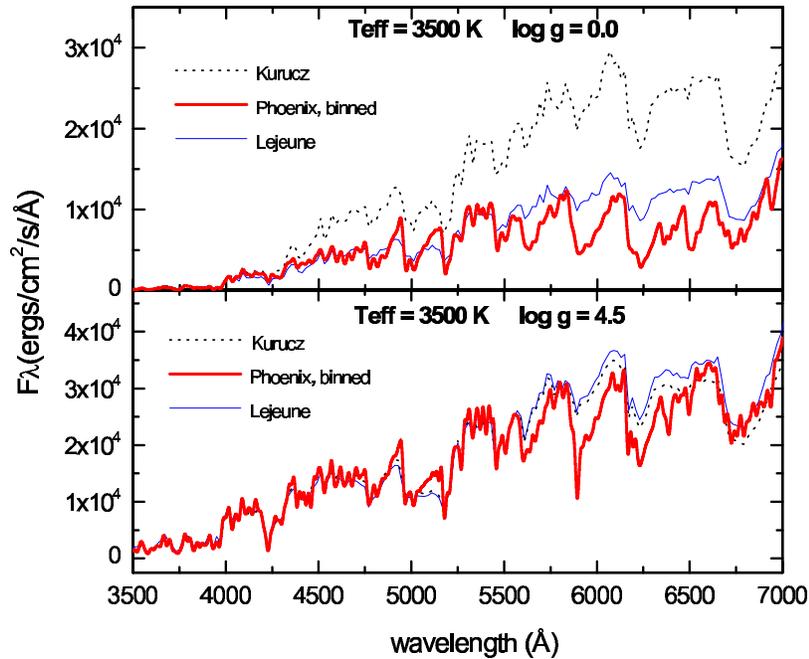}
  \caption{Comparison of the fluxes predicted by the models of Kurucz (1993), Lejeune et al. (1998), and Hauschildt et al. (1999). The latter are based on Phoenix atmospheres and predict the overall shape in a self-consistent manner. Upper panel: supergiant; lower panel: dwarf (Martins et al. 2004).}
\end{figure}

Current theoretical efforts are reflected in the synthetic library of Martins et al. (2004) who computed $\sim$1600 high-resolution stellar spectra, with a sampling of 0.3~\AA\ and covering the wavelength range from 3000 to 7000~\AA. The library was computed with the latest improvements in stellar atmospheres, incorporating non-LTE
line-blanketed TLUSTY models (Lanz \& Hubeny 2003) for hot, massive stars and spherical line-blanketed Phoenix models accounting for tri-atomic molecules (Hauschildt et al. 1999) 
for cool stars. The grid covers the full HRD for chemical abundances of twice solar, solar, half solar, and 1/10 solar. The replacement of the traditional Kurucz (1993) atmospheres at the hot and cool end is significant. Important age diagnostics such as the helium lines are strongly affected by non-LTE effects in O stars. LTE models predict too weak equivalent widths.  An example of the corresponding improvement at low $T_{\rm{eff}}$ is shown in Fig.~1. Phoenix models account for the overall energy distribution in a self-consistent way because of their extensive molecular line list. These molecular transitions are missing in the widely used Kurucz models. Lejeune et al. (1998) provided an empirical correction to the Kurucz models. This correction becomes obsolete with the new fully blanketed models.

Combining this (or any other suitable) library with evolution models allows the synthesis of any desired spectrophotometric quantity, a concept that dates back to Tinsley (1968). In the next section I will frequently rely on this technique for an age-dating of single stellar populations (SSP).

\section{Age-Dating Methods Applied to a SSP}

SSPs are defined as stellar populations whose star formation duration is short in comparison with the lifetimes of its most massive members. They are considered the simplest stellar systems and the fundamental building blocks of more complex starburst galaxies. While SSPs are idealized entities, Galactic OB associations (Massey 2003), giant H~II regions (Gonz\'alez Delgado \& P\'erez 2000), and super star clusters (SSC; Origlia et al. 2001) have been demonstrated to resemble SSPs rather closely. By testing the stellar clocks in the simplest of all laboratories, we can identify the most suitable age tracers in the absence of secondary parameters. The evolution of the optical spectrum as predicted by the synthesis models of Gonz\'alez-Delgado et al. (2004) is reproduced in Fig.~2. These synthetic population spectra make use of the library of Martins et al. (2004) introduced before. The purpose of this figure is to highlight similarities and differences between two widely used evolution models. While there is overall agreement, noticeable differences exist around 10 Myr and after 1~Gyr when RSGs and red giants are important, respectively. V\'azquez \& Leitherer (2004) discuss the sources of these differences. In the following, I will briefly highlight useful age discriminants as they appear over the first $\sim$100~Myr in the evolution of a SSP. 

\begin{figure}
  \includegraphics[height=1.0\columnwidth,angle=-90]{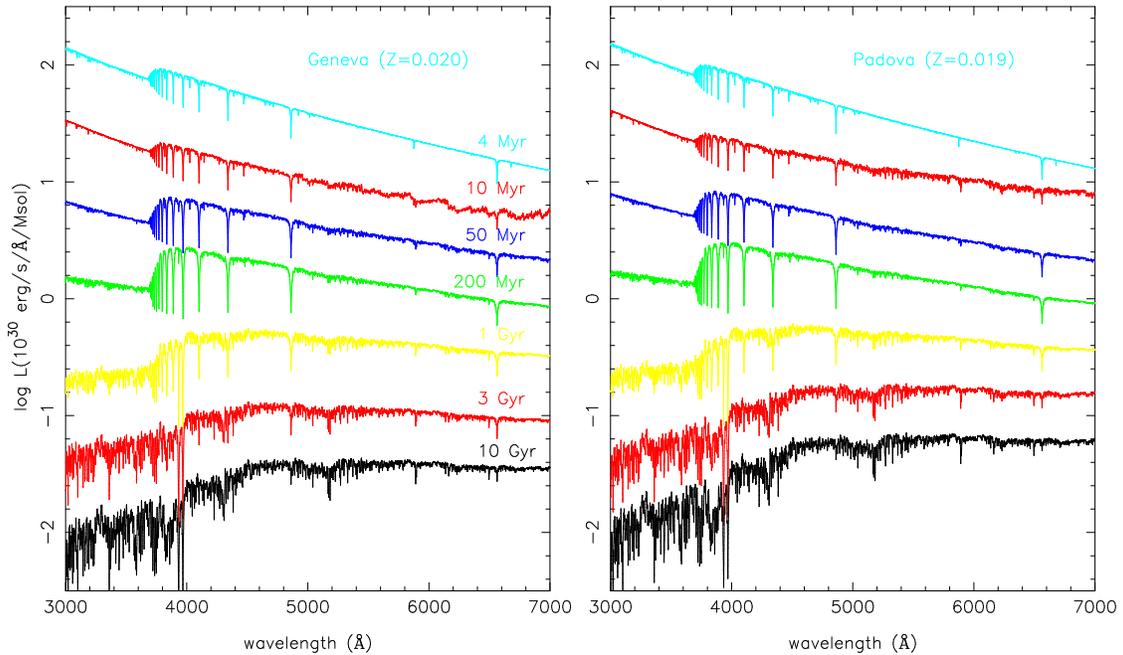}
  \caption{Theoretical spectral evolution of a SSP with solar composition predicted by the Geneva (Schaller et al. 1992; left) and the Padova (Girardi et al. 2000; right) models. Ages from top to bottom are 4 Myr, 10 Myr, 50 Myr, 200 Myr, 1 Gyr, 3 Gyr, and 10 Gyr (Gonz\'alez-Delgado et al. 2004).}
\end{figure}

{\bf 1 Myr: obscured star formation}. The first 1 to 2~Myr after the stellar birth are inaccessible to detailed age-dating because of dust obscuration. The youngest, optically identified Galactic OB associations have ages $> 1 - 2$~Myr (Massey et al. 1995), which is generally interpreted as evidence for strong obscuration at earlier ages. A ``radio supernebula'' in the nearby dwarf irregular NGC 5253 may be a starburst caught close to the act of formation (Turner \& Beck 2004). Images of the 7 mm free-free emission reveal structure in the nebula, which has a 1 pc core requiring the excitation of $\sim$10$^3$ O7 stars. The total ionizing flux within the central 20 pc is $7 \times 10^{52}$~s$^{-1}$, corresponding to 7000 O7 stars. This cluster is the dominant ionizing source in NGC~5253, yet it is totally obscured at optical wavelengths.

{\bf 4 Myr: UV stellar-wind lines}. The wavelength region between 1200~\AA\ and 2000~\AA\ is dominated by 
stellar-wind lines of, e.g., \civ\ and \siiv, the strongest
features of hot stars in a young population (e.g., Leitherer et al. 1995; 2001). In contrast, the optical and IR spectral 
regions show few, if any, spectral signatures of
hot stars, both due to blending by nebular emission and the general
weakness of hot-star features longward of 3000~\AA. Hot-star winds are radiatively driven,
with radiative momentum being transferred into kinetic momentum via absorption
in metal lines, like those observed in the satellite-UV. Since the stellar far-UV radiation field softens with time 
for an evolving SSP, the wind strength decreases, and the lines
change from being P~Cygni wind profiles during the first few Myr to
purely photospheric absorption lines after about 10~Myr. This is illustrated
in Fig.~3, which compares the UV spectra of two SSCs in the Antennae (NGC~4038/39) to a theoretically calculated age sequence. At early age, C~IV and Si~IV display P~Cygni profiles, followed by a gradual weakening due to the changing wind density and
ionization conditions. An important application of this method is at redshift $> 2$, where the restframe-UV
becomes accessible to large ground-based
telescopes (Pettini et al. 2000). 

\begin{figure}
  \includegraphics[height=0.6\columnwidth]{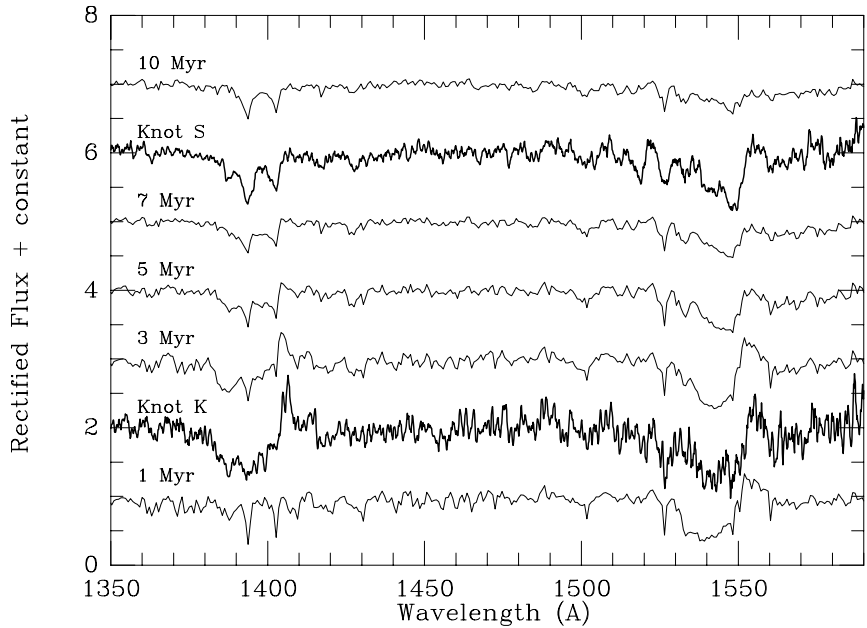}
  \caption{Comparison between the observed UV spectra of two SSCs in the Antennae (thick lines) and model spectra (thin lines) based on evolutionary population synthesis from Starburst99 (Whitmore et al. 1999). }
\end{figure}

{\bf 4 Myr: nebular emission lines}. Young stellar systems are embedded in gas. If O, B, and W-R stars are present,
the gas will be ionized and excited.  Analysis of electron-temperature 
sensitive lines in the optical and fine-structure lines in the mid-IR allows an independent determination of the stellar
heating, and therefore of the stellar far-UV radiation field and its evolution
with time. The age range for which this technique is sensitive coincides with the
evolutionary timescale of O stars, i.e., $\sim$10~Myr for a SSP.  
Evolutionary synthesis calculations coupled to photoionization models predict
strong variations of the far-UV radiation field, in particular when hot stars
with strong winds appear (e.g., Smith et al. 2002). Different models seem to converge in
their predictions and with observations longward of the He$^\circ$ edge, but
discrepancies exist around and below the He$^+$ edge. The equivalent widths of the strongest hydrogen recombination lines, such as \Ha, \Hb, or \Brg\ can be very powerful age indicators (Stasi\'nska \& Leitherer 1996) because they measure the ratio of the young, ionizing over the old, non-ionizing population. In practice, their interpretation is complicated by reddening affecting the nebular emission and the stellar continuum differently and by underlying older populations diluting the continuum.

{\bf 10 Myr: near-IR features from RGSs}. After about 5~Myr, the most massive stars of a single population evolve toward
cooler temperatures forming RSGs. The RSG continuum and lines will dominate
the red and near-IR for the following tens of Myr. Since RSG wind densities
and surface gravities are orders of magnitude lower than in their hot 
evolutionary
progenitors, narrow ($\sim$2~\kms) photospheric lines and the CO bandhead at 2.3 $\mu$m can be observed in
individual stars. In star clusters, these lines allow a velocity
dispersion measurement, and therefore a determination of the cluster mass
(Lehnert, this conference). Most RSG features are quite age sensitive because of their temperature dependence, and their anticorrelation with nebular emission lines permits studies of age gradients and propagating star formation among H~II regions (Ryder et al. 2001). Unfortunately, these lines often exhibit a strong gravity dependence as well. Therefore,
contributions from less luminous --- but more numerous --- giant stars must
be carefully evaluated in more complex populations (Rhoads 1998). 
The shift in wavelength of age-sensitive lines from the UV to the IR that
occurs with the appearance of RSGs has an added benefit: obscured clusters  can be
observed with attenuation factors reduced by an order of magnitude 
relative to the V band.

{\bf 25 Myr: isochrone analysis of B stars}. The color-magnitude diagram (CMD) is the tool of choice for studying the star-formation history systems which are resolved into stars and whose colors are age sensitive (the latter is not the case for O stars). Deep CMDs contain stars born over the whole mass range and can be used to derive ages from, e.g., the main-sequence turn-off (Aparicio \& Gallart 2004). Given the information density in an observed CMD, an isochrone analysis is powerful enough to provide not only the age of the population but at the same time also the chemical composition, the cluster mass, and the stellar initial mass function). This method relates the number of stars populating different regions of the observational
CMD to the density distribution of stars in the CMD as a function of age, mass, and
metallicity as predicted by the stellar evolution theory. NGC 330, a rich cluster in the LMC, has become an often used test case for models (e.g., Keller et al. 2000). It has a well-defined turn-off dominated by B stars, as well as a significant RSG population which provides constraints on the evolved population. Its turn-off age is $25 \pm 5$~Myr. NGC~330 serves as an example that nature can be more complex than standard evolution models: rapid rotators have been identified to produce a population of overluminous B stars contaminating the main-sequence turn-off (Grebel et al. 1996). While these stars can easily be accounted for in the CMD, they would skew the age estimate towards younger ages if integrated photometry or spectroscopy of the cluster were taken.

{\bf 25 Myr: shock-sensitive lines}. A few percent of the total radiative stellar luminosity of a young population
is converted into mechanical luminosity via stellar winds and supernovae.
Integrated over their lifetimes, the energy release by stellar winds is
comparable to that of a supernova event: Mass-loss rates of about
$10^{-5}$~\Myr\ and wind velocities on the order of
$10^3$~\kms\ result in mechanical energies around $10^{51}$~erg over $10^7$~yr (Leitherer et al. 1999).
The non-thermal energy input is traced, e.g., by 
[Fe~II] $\lambda$1.26~$\mu$m and $\lambda$1.64~$\mu$m, which are often observed in evolved
starbursts and can be used to age-date  population
(e.g., Calzetti 1997; Dale et al. 2004). Under normal conditions, Fe is a refractory element locked in grains and therefore highly depleted. The appearance of strong [Fe~II] lines can be understood in terms of
an increased gas-phase abundance of Fe due to grain destruction in shocks (Forbes \& Ward 1993).

{\bf 30 Myr: Balmer absorption lines}. The spectra of early-type stars are characterized by strong hydrogen Balmer and neutral helium absorption lines (Walborn \& Fitzpatrick 1990). However, the detection of these stellar features at optical wavelengths in the spectra of young starbursts is difficult because H and He~I absorption features are coincident with the nebular emission lines that mask the absorption features. Once the starburst is old enough for ionizing radiation to be negligible, the stellar H and He absorption lines become a useful diagnostic. The strength of the Balmer absorption lines is strongly age dependent. The equivalent widths increase with age until about a few hundred Myr when the stellar population is dominated by early-A type stars; they range between 2 and 16~\AA\ if the cluster is formed at solar metallicity. An application of this method is in Fig.~4 where the synthetic models of Gonz\'alez Delgado et al. (2004) are used to determine ages of star clusters in the LMC.  The advantages of using Balmer and He I lines in absorption to date starbursts with respect to using nebular emission lines are twofold: (i) the age can be constrained over a much wider range, and (ii) the strengths of the absorption lines are not affected by extinction (but see the following section) or by the leaking of ionizing photons.

\begin{figure}
  \includegraphics[height=0.7\columnwidth,angle=-90]{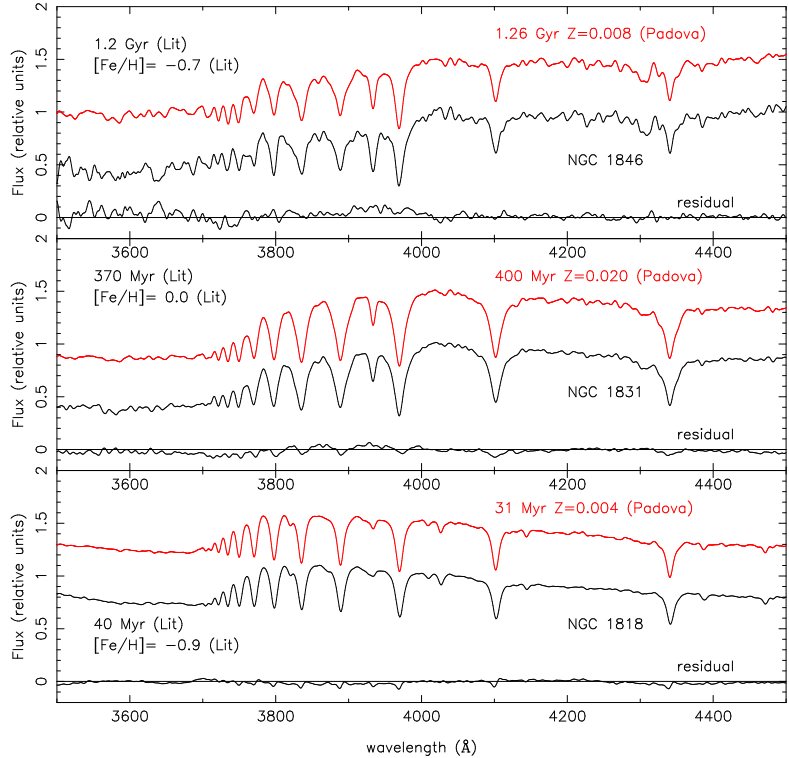}
  \caption{Intermediate-resolution spectra of the three LMC clusters NGC 1818 (bottom panel), NGC 1831 (center), and NGC~1846 (top) compared with theoretical spectra computed by Gonz\'alez Delgado et al. (2004). The models are the upper spectra in each panel. The cluster ages are literature values, whereas the model ages are derived from the best-fit model.}
\end{figure}

{\bf 30 Myr: optical and near-IR colors}. Colors are the prime age indicators in intermediate and old stellar populations. Reddening and the optical color degeneracy of hot stars limit their use to
ages of more than about 30~Myr. There is very little metallicity dependence
in young systems due to the lack of strong metal lines in the optical and
near-IR. Metallicity does enter via evolution models when RSGs affect the
colors (between 10 and 50~Myr). However, as I will discuss below, this particular model prediction is often incorrect due to wrong RSG parameters in stellar evolution models for metal-poor populations.
The main challenge in interpreting colors of young systems is to break the
{\em age-reddening degeneracy}. The reddening vector is parallel to the 
isochrones in most color-color diagrams, and the reddening correction
uncertainty is often much larger than the desired precision of the age
determination. Reddening corrections are not only an issue
for interpreting colors but also for recombination-line equivalent widths (e.g., \Hb),
as the obscuration of stars and gas is not the same in a young population (Calzetti 2001).

\section{Degeneracies in Complex Systems} 

Starburst galaxies are a complex superposition of individual starburst regions, embedded in gas and dust having highly irregular morphologies. Reading off the age-dating clock becomes far from trivial in such environments. A case in point is the center of the nearby starburst galaxy M83, as reproduced in Fig.~5 (Harris et al. 2001). Located within the main bar of M83, the center of the galaxy contains a well-defined nuclear subsystem.  The optically bright nucleus
is offset from the dynamical center opposite to the
starburst ring. Dense molecular gas has been detected and is generally
concentrated to the north of the starburst ring. The optically visible nucleus, as well
as a second dust hidden nucleus, are dominated by
evolved red giant stars. These observations suggest a complicated morphology, where starbursts have occurred in multiple bursts and in an inhomogeneous interstellar medium (ISM). Age-dating methods applied to such systems suffer from manifold {\em age-metallicity-IMF-ISM degeneracies}.

The {\em age-metallicity degeneracy}, while a major concern at old ages (Trager et al. 2000), is less of an issue for the age range relevant to starbursts. Metallicity effects in hot-star atmospheres are relatively minor because of reduced line blanketing at high \Teff. The chemical composition does of course enter via the metal-dependent stellar evolution. However, any self-enrichment is generally small since the nucleosynthetic products of massive stars appear to require much longer than $10^7$~yr to mix with the surrounding ISM (Kobulnicky \& Skillman 1997). In any case, independent techniques are available for estimating the metallicity in young starburst, such as emission-line spectroscopy of H~II regions.

Most age-dating methods rely on evaluating the luminosity weighted light contributions from stars having different ages. Since there exists an $L \propto M$ relation, mass can always be traded for luminosity, and an {\em age-IMF degeneracy} is introduced. Therefore, IMF variations can mimic age gradients. While the evidence for significant IMF variations in local star clusters is weak (Kroupa 2002), such variations cannot be fully excluded in the dense starburst environment. Figer et al. (1999) obtained a determination of the upper IMF for the Arches and Quintuplet clusters, two extraordinary young clusters near the Galactic center. They found an IMF slope that is significantly flatter than the average for young clusters elsewhere in the Galaxy. If IMF variability is a concern, combining spectral features that respond differently to age and IMF can help overcome the age-IMF degeneracy. In general, age indicators tracing stars in the red part of the HRD are less IMF- and more age-sensitive than their counterparts in the blue. There are two reasons for this behavior. (i) Since the most luminous blue stars are more luminous than the most luminous red stars, RSGs have descended from the main-sequence where the initial $M-L$ relation is steeper than for more massive blue stars. As a result, RSGs sample a smaller mass range in the integrated spectrum and are less IMF sensitive. (ii) The evolutionary tracks are less degenerate in the red than in the blue. A particular HRD location in the blue may be occupied stars having vastly different {\em current} masses, whereas the mass distribution in the red is quite homogeneous. Taken together, both effects suppress the IMF sensitivity of RSG related spectral features.  

\begin{figure}
  \includegraphics[height=0.7\columnwidth]{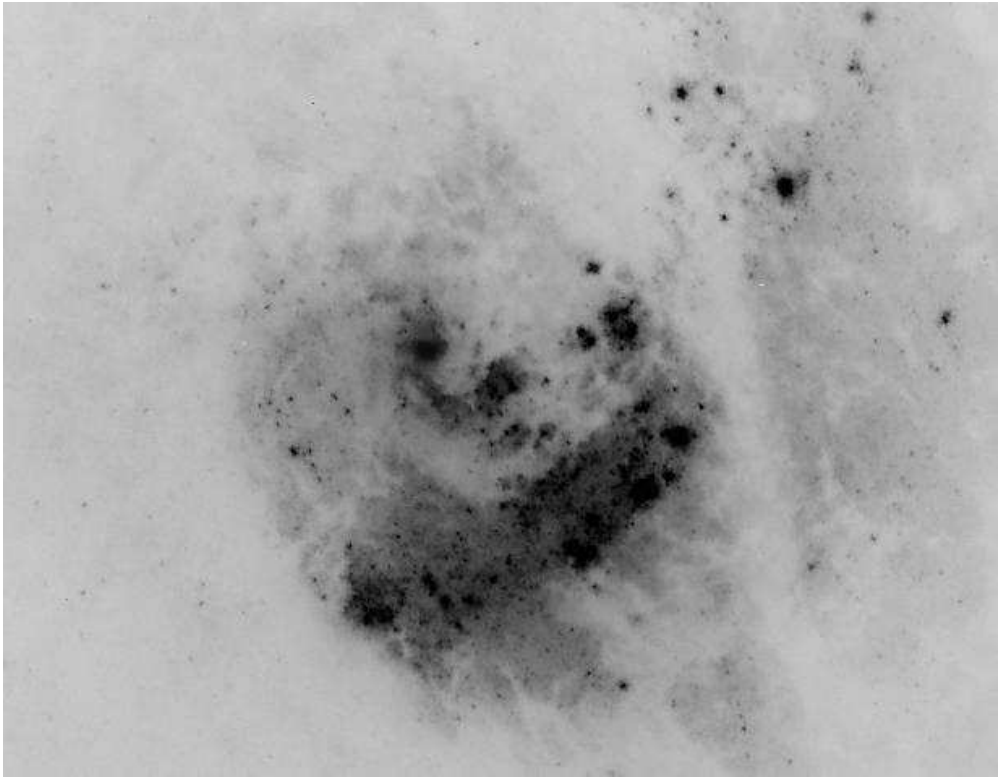}
  \caption{Multicolor image of the center of M83 constructed from HST/WFPC2 F300W, F547M, and F814W images. North is at the top, and east is at the left. Field size is $36'' \times 36''$, corresponding to 650~pc $\times$ 650~pc (Harris et al. 2001).}
\end{figure}

Breaking the {\em age-ISM degeneracy} can be the most daunting challenge, as the following example of starbursts containing W-R stars demonstrates. W-R galaxies are a subset of the starburst class, whose 
observational characteristic is the broad emission bump around 4640 -- 4690~\AA\ (Conti 1991). The compilation 
by Schaerer et al. (1999) lists 139 members. 

W-R galaxies are important because they permit age determinations via {\em stellar} spectral 
features, as opposed to indirect tracers based on gas and/or dust emission. Hot stars are notoriously elusive even 
in the strongest starbursts because their spectral signatures are too weak, coincide with nebular emission lines, 
or are in the satellite-UV. W-R stars are the only hot, massive stellar species detectable at optical wavelengths. This is because they have the strongest stellar winds, which in combination with their high temperatures produce broad ($\sim$1000~km~s$^{-1}$) emission lines not coinciding with emission from H~II regions. Examples are N~III $\lambda$4640, C~III $\lambda$4650, He~II $\lambda$4686, C~III $\lambda$5696, and C~IV $\lambda$5808. The mere detection of such features proves the presence of stars with masses above 40 -- 60~M$_\odot$ and ages of 2~--~6~Myr (depending on chemical composition) since only stars this massive can evolve into the W-R phase. This powerful diagnostic can be used for, e.g., inferring a massive star population when the space-UV is inaccessible due to dust (IR-luminous galaxies), or when broad nebular lines veil the O stars (Seyfert2 galaxies). 

We are currently involved in an optical+near-IR survey of luminous starburst galaxies (Le\~ao, Leitherer, \& Bresolin, in prep.). Our goal is to investigate the 
stellar content from purely stellar tracers. The starburst galaxies are drawn from an IRAS sample of Lehnert \& Heckman (1995). The galaxies have $10 < \log L_{\rm{IR}}/$L$_\odot < 11.5$, warm IR colors, and are not AGN dominated. 
Detection of the W-R features, together with other standard diagnostics allows 
us to probe the age and IMF. In particular, we can search for or against evidence of a peculiar IMF at high metallicity, as indicated, e.g., by IR observations.

Thornley et al. (2000) carried out an ISO spectroscopic
survey of 27 starburst galaxies 
with a range of luminosities from $10^8$ to $10^{12}$~L$_\odot$. 
The [Ne~III] 15.6~$\mu$m and [Ne~II] 12.8~$\mu$m lines are
particularly useful. The ionization
 potentials of neutral and ionized Ne are 22 and 41~eV, respectively.
The two lines are very close in wavelength (12.8 and 15.6~$\mu$m), 
and they have similar critical densities. This makes the line ratio a sensitive
probe for various star formation parameters, in particular the upper mass 
cutoff of the IMF ($M_{\rm{up}}$) and the age and duration of the starburst.
Photoionization models of Thornley et al. and Rigby \& Rieke (2004) suggest
that stars more 
massive than about 35 to 40~M$_\odot$ are deficient in the observed sample. 
Either they never formed because of a peculiar IMF, or they have already 
disappeared due to aging effects. This result echos that obtained
from ground-based near-IR spectroscopy: the strategic recombination lines
are powered by a soft radiation field originating from stars less massive than
$\sim$40~M$_\odot$ (Doyon et al. 1992).
An upper-mass cutoff as low as 40~M$_\odot$, however, is difficult to reconcile with 
the ubiquitous evidence of very massive stars with
masses of up to 100~M$_\odot$ in many starburst regions (e.g., 
Leitherer et al. 1996; Massey \&
Hunter 1998; Gonz\'alez Delgado et al. 2002). 
Therefore the alternative explanation,
an aged starburst seems more plausible. Under this assumption, stars of 
masses 50~--~100~M$_\odot$ are initially formed in most galaxies, but the
starbursts are observed at an epoch when these stars are no longer present.
This implies that the inferred burst durations must be less than a few Myr. Such short burst timescales are surprising,
in particular for luminous, starburst galaxies whose dynamical timescales
can exceed tens of Myr. Both the peculiar IMF or the short starburst timescales in dusty,
IR-bright starbursts are quite unexpected and pose a challenge to 
conventional models in which the starburst is fed by gas inflow to
the nucleus over tens of Myr as a result of angular momentum loss. 

Early results from our W-R survey urge caution when relating nebular IMF tracers to the actual stellar content. Several metal-rich starburst galaxies exhibiting a soft radiation field do in fact have a substantial \mbox{W-R} population. Examples are the archetypal starburst galaxies NGC~1614, NGC~2798, or NGC~3690. We clearly detect the tell-tale signatures of W-R stars in these galaxies. Unless stellar evolution proceeds differently than in our Galaxy, massive O stars must be present as well. The fact that we do not {\em directly} observe these O stars is no contradiction: their spectral lines will be hidden by coinciding nebular emission lines. We should, however, detect their ionizing radiation in a proportion predicted by the measured number of W-R stars and the expected ratio of W-R/O stars. The deficit of radiation suggests that indirect age and IMF tracers, like nebular lines, still
require careful calibration, in particular when applied to dusty,
metal-rich starbursts.

Interpretation of the stellar W-R feature itself may sometimes be complicated by the inhomogeneous structure of the surrounding ISM. As part of a larger project to quantify the stellar and interstellar properties of local galaxies undergoing active star formation (Chandar et al. 2004), we have obtained HST STIS long-slit far- and near-UV spectra for 15 local starburst galaxies. The target galaxies were selected to cover a broad range of morphologies, chemical composition, and luminosity. Some of them are known W-R galaxies. The UV counterpart of the optical He~II $\lambda$4686 line is the 3 $\rightarrow$ 2 transition of He$^+$ at 1640~\AA.
  Broad He~II $\lambda$1640 emission is seen in the UV spectra of individual Galactic and Magellanic Cloud W-R {\em stars} (e.g., Conti \& Morris 1990) but is not prevalent in the integrated spectra of {\em galaxies} because of the overwhelming light contribution from OB stars in the UV.

NGC 3125-1 has by far the largest He~II $\lambda$1640 equivalent width in the sample. The 1640~\AA\ emission is quite prominent and appears significantly stronger than the C~IV emission at 1550~\AA. In addition to this feature, N~IV $\lambda$1488 and N~IV $\lambda$1720 emission are also detected, consistent with the interpretation that the strong He~II $\lambda$1640 emission arises in the winds of massive stars. 6100 WNL stars are estimated to reside in this starburst region, making it the most W-R-rich known example of an individual starburst cluster in the local universe. What makes this region extraordinary is the number of W-R stars relative to other massive stars. The W-R features are almost undiluted compared with those seen in single W-R stars. Therefore, most of the continuum light must be emitted by the very same W-R stars. Quantitative modeling leads to approximately equal W-R and O-star numbers. Such an extreme ratio is excluded by stellar evolution models, even for a 3~--~5~Myr old starburst with a most unusual IMF.

  How can we reconcile these observations with our current understanding of W-R stars? The UV continuum slope is normally dominated by OB stars; however, in this case the large number of W-R stars also makes a significant contribution. One possible explanation is that we are seeing the W-R stars through a ``hole'' where their energetic winds have blown out the natal cocoon earlier than have the OB stars. The often large reddening derived in starburst regions is consistent with the generally accepted scenario in which very young clusters remain embedded in their natal material until energetic stellar winds from evolving massive stars blow out the surrounding gas and dust. 

     If W-R stars are preferentially less attenuated than OB stars in \mbox{NGC 3125-1}, the {\em equivalent width} of He~II $\lambda$1640 and other W-R lines is skewed towards artificially large values because for a standard stellar population surrounded by a homogeneous ISM, the continuum is emitted by the less massive OB stars. If the ISM is inhomogeneous, {\em the equivalent widths of purely stellar lines and the associated age determination become reddening dependent}.

\section{Trouble Spot: Red Supergiants}

Despite the challenges of correctly interpreting astrophysical clocks in complex starbursts, the {\em stellar} modeling is rather reliable over most of the HRD (see, e.g., the comprehensive review of Bruzual 2003). One remaining nagging issue concerns RSGs. I will elaborate on this trouble spot using blue compact dwarfs (BCDs) as an example. BCDs are thought to be strongly starbursting galaxies powered by
ionizing stars and with an underlying population of red stars whose precise age is still under discussion
(e.g., Aloisi, Tosi, \& Greggio 1999; Thuan, Izotov, \& Foltz 1999). There is, however,
convincing evidence for a significant number of RSGs in those objects with well-established
CMDs (e.g., Schulte-Ladbeck et al. 2001). 

\begin{figure}
  \includegraphics[height=0.7\columnwidth,angle=0]{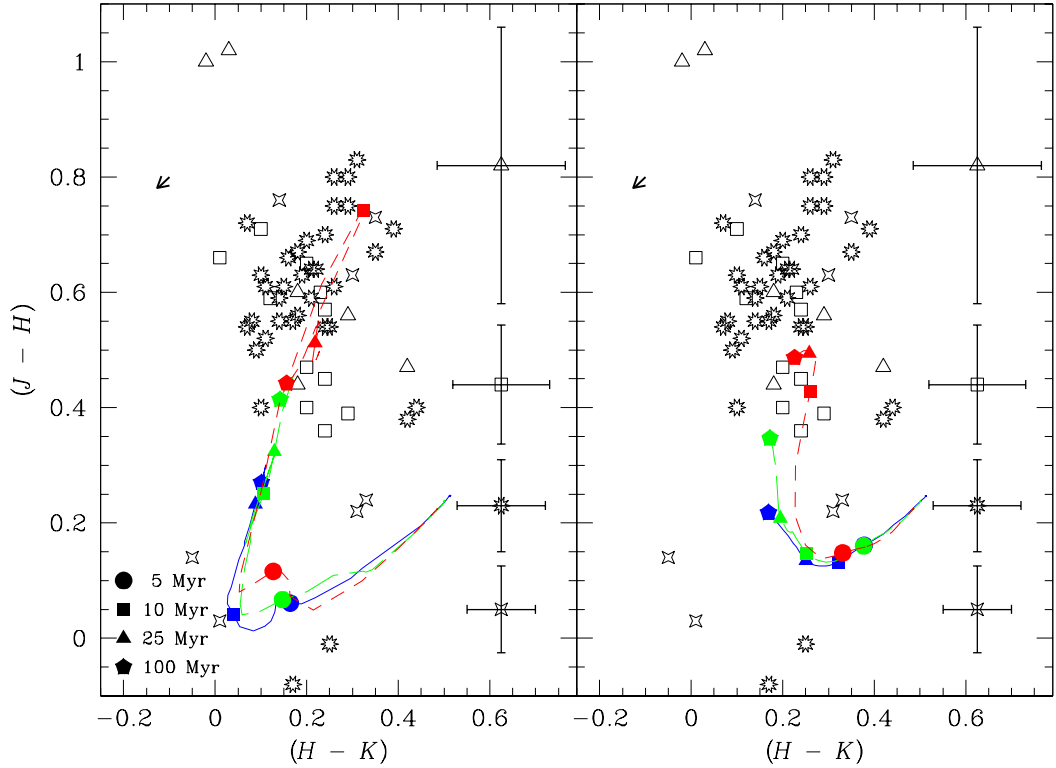}
  \caption{Color-color diagrams of four samples of BCDs compared with starburst models of
$Z = 0.001$ (solid line), 0.004 (long-dashed), and 0.02 (short-dashed
line) based on the Geneva tracks. Filled symbols indicate the ages of the models. Open symbols
show the data: triangles (Doublier et al. 2001), squares (Noeske et al. 2003), starred circles
(Thuan 1983), and four-legged stars (Telles 2004). The error bars are the dispersion value in each
sample. The solid vector indicates the reddening correction for $A_{\rm{V}} = 0.25$. Left: models for a
SSP; right: continuous star formation (V\'azquez \& Leitherer 2004).}
\end{figure}

V\'azquez \& Leitherer (2004) collected near-IR photometry of a sample of BCDs from the literature and compared them to synthesis models. The large-aperture photometry is
sensitive to the young starburst, the surrounding field consisting of potential earlier
starburst episodes, and the older underlying population. Therefore any RSGs present in
these galaxies will affect or even dominate the photometry.
The observations collected from the literature are plotted and compared to models in Fig.~6. Both reddening and line emission are negligible at IR wavelengths. The synthetic colors predicted for a SSP  and for continuous star formation are in the left and right panels of Fig.~6, respectively. 
Fig.~6 (left) suggests reasonable agreement between the bulk of the data and the computed
colors of a SSP with solar chemical composition.

The synthetic models for SSPs were terminated at an age of 100~Myr. Higher ages are unrealistic, as BCDs are defined via their emission lines and blue colors. 100 Myr old starbursts would not be classified as BCDs anymore. The details of the  star formation during the first tens of Myr, however, is a subject of debate. Dwarf galaxies are known to have had rather complex star-formation
histories, with periods of quiescence and intermittent bursts of star formation (e.g., Greggio
et al. 1998). Depending on the burst frequency, the effective star formation may mimic a steady-state situation. This is addressed in the right panel of Fig.~6.
As expected, the imprint of the RSGs is more diluted, and the colors are more degenerate
than for a SSP. If the star-formation history in BCDs were constant,
the comparison between the data and the models would force one to postulate ages far
in excess of 100 Myr. While this would not necessarily be in conflict with observational
selections (ionizing stars are continuously replenished), the associated gas consumption
would become entirely unreasonable. BCDs do not constantly form stars over 1 Gyr.
Therefore the appropriate star-formation scenario is between the extremes plotted
in Fig.~6, but most likely much closer to the SSP case in the left panel.

Does this suggest consistency between the observed and synthetic colors? The only
track in Fig.~6 (left) that matches the data points is the one at solar chemical composition.
The other tracks at lower abundance are significantly bluer and fail to reproduce the
observed colors regardless of the assumed age and reddening correction. Only the solar
models produce RSGs in large enough numbers and with sufficiently low \Teff\ to reach the colors covered by the data points. Yet, the approximate average oxygen abundance of the sample is 20\% solar. Therefore the assumption of solar composition is invalid, and the applicable models are those with
$Z = 0.004$. Once forced to compare the data to the $Z = 0.004$ tracks, one arrives at
the inescapable conclusion that the predicted colors of evolutionary models for
metal-poor populations with a significant RSG component are incorrect. This conclusion is
unchanged for either the Padova or Geneva tracks.
Our results echo those of Massey \& Olsen (2003), who demonstrated that both the
Geneva and Padova evolution models fail to predict the location of RSGs and the Large
and Small Magellanic Clouds. The evidence of failure at solar chemical composition is
much weaker, if present at all. 

\section{Back to the Future}

Tinsley (1977) characterized the dawn of the era of quantitative age-dating: ``{\em The colors and spectra of galaxies are therefore sensitive to the presence of OB stars, red giants of the old-disk type, and/or metal-poor giants of the halo type. From their relative numbers comes information on the age and metallicity distribution of stars in different regions of galaxies, which is turn gives clues about galactic evolution. Despite decades of observational and theoretical work in this field, many interesting details are still out of reach..... A challenge for the future is to exploit further what can be seen in the integrated light, of systems whose HRDs cannot be drawn point by point, to obtain a fuller understanding of the evolution of stars and stellar systems.}'' The field pioneered by her has gone a long way since these early days.

\begin{theacknowledgments}
  I wish to thank the organizers for providing travel support and for making this meeting so successful.

\end{theacknowledgments}

{}


\begin{thebibliography}{}

\bibitem{} Aloisi, A., Tosi, M., \& Greggio, L. 1999, AJ, 118, 302

\bibitem{} Aparicio, A., \& Gallart, C. 2004, AJ, in press

\bibitem{} Bonanos, A. Z., Stanek, K. Z., Udalski, A., Wyrzykowski, L., Zebrun, K., Kubiak, M., Szymanski, M. K., Szewczyk, O., Pietrzynski, G., \& Soszynski, I. 2004, ApJ, 611, L33

\bibitem{} Bruzual, G. 2003, in Galaxies at High Redshift, XI Canary Islands Winter School of Astrophysics, ed. 
      I. P\'erez-Fournon, M. Balcells, F. Moreno-Insertis \& F. S\'anchez (Cambridge: CUP), 185

\bibitem{} Calzetti, D. 1997, AJ, 113, 162

\bibitem{} ------. 2001, PASP, 113, 1449

\bibitem{} Chandar, R., Leitherer, C., \& Tremonti, C. A. 2004, ApJ, 604, 153

\bibitem{} Chiosi, C., \& Maeder, A. 1986, ARAA, 24, 329

\bibitem{} Conti, P. 1991, ApJ, 377, 115

\bibitem{} Conti, P., \& Morris, P. W. 1990, AJ, 99, 898

\bibitem{} Dale, D. A., Roussel, H., Contursi, A., Helou, G., Dinerstein, H. L., Hunter, D. A., Hollenbach, D. J., Egami, E., Matthews, K., Murphy, T. W., Jr., Lafon, C. E., \& Rubin, R. H. 2004, ApJ, 601, 813 

\bibitem{} Doublier, V., Caulet, A., \& Comte, G. 2001, A\&A, 367, 33

\bibitem{} Doyon, R., Puxley, P. J., \& Joseph, R. D. 1992, ApJ, 397, 117

\bibitem{} Figer, D. F., Kim, S. S., Morris, M., Serabyn, E., Rich, R. M., \& McLean, I. S. 1999, ApJ, 525, 750

\bibitem{} Forbes, D. A., \& Ward, M. J. 1993, ApJ, 416, 150

\bibitem{} Girardi, L., Bressan, A., Bertelli, G., \& Chiosi, C. 2000, A\&A, 141, 371

\bibitem{} Gonz\'alez Delgado, R. M., Cervi\~no, M., Martins, L. P., Leitherer, C., \& Hauschildt, P. 2004, MNRAS, in press

\bibitem{} Gonz\'alez Delgado, R. M., Leitherer, C., Stasi\'nska, G., \& Heckman, T. M. 2002,
ApJ, 580, 824

\bibitem{} Gonz\'alez Delgado, R. M., \& P\'erez, E. 2000, MNRAS, 317, 64

\bibitem{} Grebel, E. K., Roberts, W. J., \& Brandner, W. 1996, A\&A, 311, 470

\bibitem{} Greggio, L., Tosi, M., Clampin, M., de Marchi, G., Leitherer, C., Nota, A., \& Sirianni, M. 1998, ApJ, 504, 725

\bibitem{} Harris, J., Calzetti, D., Gallagher, J. S., Conselice, C. J., \& Smith, D. A. 2001, AJ, 122, 3046

\bibitem{} Hauschildt, P. H., Allard, F., \& Baron, E. 1999, ApJ, 512, 377

\bibitem{} Keller, S. C., Bessell, M. S., \& Da Costa, G. S. 2000, AJ, 119, 1748

\bibitem{} Kobulnicky, H. A., \& Skillman, E. D. 1997, ApJ, 489, 636 

\bibitem{} Kroupa, P. 2002, in Modes of Star Formation and the Origin of Field Populations, ed. E. K. Grebel \& W. Brandner (San Francisco: ASP), 86

\bibitem{} Kudritzki, R. P., \& Puls, J. 2000, ARAA, 38, 613

\bibitem{} Kurucz, R. 1993, ATLAS9 Stellar Atmosphere Programs and 2 km/s grid. Kurucz CD-ROM No. 13 (Cambridge: SAO)

\bibitem{} Lanz, T., \& Hubeny, I. 2003, ApJS, 146, 417

\bibitem{} Lehnert, M. D., \& Heckman, T. M. 1995, ApJS, 97, 89 

\bibitem{} Leitherer, C., Le\~ao, J. R. S., Heckman, T. M., Lennon, D. J., Pettini, M., \& Robert, C. 2001, ApJ, 550, 724

\bibitem{} Leitherer, C., Robert, C., \& Heckman, T. M. 1995, ApJS, 99, 173

\bibitem{} Leitherer, C., Schaerer, D., Goldader, J. D., Gonz\'alez Delgado, R. M., Robert, C., Foo Kune,
    D., de Mello, D., Devost, D., \& Heckman, T. M. 1999, ApJS, 123, 3

\bibitem{} Leitherer, C., Vacca, W. D., Conti, P. S., Filippenko, A. V., 
    Robert, C., \& Sargent, W. L. W. 1996, ApJ, 465, 717

\bibitem{} Lejeune, T., Cuisinier, F., \& Buser, R. 1998, A\&A, 130, 65 

\bibitem{} Lejeune, T., \& Fernandes, J. 2002, Observed HR Diagrams and Stellar Evolution, (San Francisco: ASP) 

\bibitem{} Maeder, A., \& Meynet, G. 2000, ARAA, 38, 143

\bibitem{} Martins, L. P., Gonz\'alez Delgado, R. M., Leitherer, C., Cervi\~no, M., \& Hauschildt, P. 2004, MNRAS, in press

\bibitem{} Massey, P. 2003, ARAA, 41, 15

\bibitem{} Massey, P., \& Hunter, D. A. 1998, ApJ, 493, 180

\bibitem{} Massey, P., Johnson, K. E., \& Degioia-Eastwood, K. 1995, ApJ, 454, 151

\bibitem{} Massey, P. \& Olsen, K. A. G. 2003, AJ, 126, 2867

\bibitem{}  Mehlert, D., Noll, S., Appenzeller, I., Saglia, R. P., Bender, R., B\"ohm, A., Drory, N., Fricke, K., Gabasch, A., Heidt, J., Hopp, U., J\"ager, K., M\"ollenhoff, C., Seitz, S., Stahl, O., \& Ziegler, B. 2002, A\&A, 393, 809 

\bibitem{} Noeske, K. G., Papaderos, P., Cair\'os, L. M., \& Fricke, K. J. 2003, A\&A, 410, 481

\bibitem{} Origlia, L., Leitherer, C., Aloisi, A., Greggio, L., \& Tosi, M. 2001, AJ, 122, 815

\bibitem{} Pettini, M., Steidel, C. C., Adelberger, K. L., Dickinson, M. \& Giavalisco, M. 2000, ApJ, 528, 96 

\bibitem{} Rauw, G., De Becker, M., Naz\'e, Y., Crowther, P. A., Gosset, E., Sana, H., van der Hucht, K. A., Vreux, J.-M., \& Williams, P. M. 2004, A\&A, 420, L9

\bibitem{} Rhoads, J. E. 1998, AJ, 115, 472

\bibitem{} Rigby, J. R., \& Rieke, G. H. 2004, ApJ, 606, 237 

\bibitem{} Ryder, S. D., Knapen, J. H., \& Takamiya, M. 2001, MNRAS, 323, 663

\bibitem{} Schaerer, D., Contini, T., \& Pindao, M.\ 1999, A\&AS, 136, 35 

\bibitem{} Schaller, G., Schaerer, D., Meynet, G., \& Maeder, A. 1992, A\&AS, 96, 269

\bibitem{} Schulte-Ladbeck, R. E., Hopp, U., Greggio, L., Crone, M. M., \& Drozdovsky, I. O. 2001, AJ, 121, 3007

\bibitem{} Smith, L. J., Norris, R. P. F., \& Crowther, P. A. 2002, MNRAS, 337, 1309

\bibitem{} Stasi\'nska, G., \& Leitherer, C. 1996, ApJS, 107, 662

\bibitem{} Telles, E. 2004, private communication

\bibitem{} Thomas, D., Maraston, C., \& Korn, A. 2004, MNRAS, 351, L19

\bibitem{} Thornley, M.~D., Schreiber, N.~M.~F., Lutz, D., Genzel, R., Spoon, H.~W.~W., Kunze, D., 
     \& Sternberg, A. 2000, ApJ, 539, 641 

\bibitem{} Thuan, T. X. 1983, ApJ, 268, 667

\bibitem{} Thuan, T. X., Izotov, Y. I., \& Foltz, C. B. 1999, ApJ, 525, 105

\bibitem{} Tinsley, B. M. 1968, ApJ, 151, 547

\bibitem{} ------. 1977, in The HR Diagram --- The 100th Anniversary of Henry Norris Russell, ed. A. G. Davis Philip \& D. S. Hayes (Dordrecht: Reidel), 247

\bibitem{} Trager, S. C., Faber, S. M., Worthey, G., \& Gonz\'alez, J. J. 2000, AJ, 119, 1645

\bibitem{} Turner, J. L., \& Beck, S. C. 2004, ApJ, 602, L85 

\bibitem{} V\'azquez, G. A., \& Leitherer, C. 2004, ApJS, in press

\bibitem{} Walborn, N. R., \& Fitzpatrick, E. L. 1990, PASP, 102, 379

\bibitem{} Whitmore, B. C., Zhang, Q., Leitherer, C., Fall, S. M., Schweizer, F., \& Miller, B. W. 1999, AJ, 118, 1551

\bibitem{} Yi, S. K. 2003, ApJ, 582, 202

\end{thebibliography}
\end{document}